# Adaptive Neuro-Fuzzy Extended Kalman Filtering for Robot Localization

**Ramazan Havangi, Mohammad Ali Nekoui and Mohammad Teshnehlab**

**Faculty of Electrical Engineering, K.N. Toosi University of Technology
Tehran, Iran**

## Abstract

Extended Kalman Filter (EKF) has been a popular approach to localization a mobile robot. However, the performance of the EKF and the quality of the estimation depends on the correct a priori knowledge of process and measurement noise covariance matrices ($Q_k$ and $R_k$, respectively). Imprecise knowledge of these statistics can cause significant degradation in performance. This paper proposed the development of an Adaptive Neuro-Fuzzy Extended Kalman Filtering (ANFEKF) for localization of robot. The Adaptive Neuro-Fuzzy attempts to estimate the elements of $Q_k$ and $R_k$ matrices of the EKF algorithm, at each sampling instant when measurement update step is carried out. The ANFIS supervises the performance of the EKF with the aim of reducing the mismatch between the theoretical and actual covariance of the innovation sequences. The free parameters of ANFIS are trained using the steepest gradient descent (SD) to minimize the differences of the actual value of the covariance of the residual with it's theoretical value as much possible. The simulation results show the effectiveness of the proposed algorithm.

## 1. Introduction

Mobile robot localization is the problem of estimating a robot pose (position, orientation) relative to its environments. Two different kinds of localization exist: relative and absolute. Relative localization is also known as dead-reckoning (DR). It is realized through the measures provided by sensors measuring the dynamics of variables internal to the vehicle. Typical internal sensors are encoders which are fixed to the axis of the driving wheels. The basic drawback of this method is that the error of robot's position and orientation generally grows unbounded with time. Absolute localization is performed processing the data provided by a proper set of sensors measuring some parameters of the environment in which the vehicle is operating. The methods to obtain absolute measurements can be divided into methods based on the use of landmarks and methods based on the use of maps. The main drawback of absolute measures is their dependence on the characteristics of the environment. Possible changes to environmental parameters may give rise to erroneous interpretation of the measures provided by the localization algorithm. In this paper, we integrate the advantages of "the relative localization" and "the absolute localization" and make them complementary, which will enable the mobile robot to localize itself more accurately. To this purpose, data provided from odometric, laser range finder and MAP are combined together through EKF. The localization based on EKF proposed in the literatures [1-9] for the estimation of robot pose. However, a significant difficulty in designing an EKF can often be traced to incomplete a priori knowledge of the process covariance matrix $Q_k$ and measurement noise covariance matrix $R_k$ [10-13]. In most robot localization application these matrices are unknown. On the other hand, it is well known how poor estimates of noise statistics may seriously degrade the Kalman filter performance [12], [16]. One of the efficient ways to overcome the above weakness is to use an adaptive algorithm for localization. There have been many investigations in the area of adaptive algorithm for robot localization [5], [8], [9], [18], [19], [20]. In [15] a Neuro-fuzzy assisted extended kalman filter-based approach for simultaneous localization and mapping (SLAM) problem is presented. In this algorithm, a Neuro-fuzzy approach is employed to adapt the matrix $R_k$ only while $Q_k$ is completely known.

Also, as the computational load of this algorithm is very high, it cannot be implemented in real-time application. In this paper the EKF coupled with adaptive Neuro-Fuzzy Inference System has been presented to adjust the matrices $Q_k$ and $R_k$. Main advantage of this algorithm, compared to prior ones, is its fast and efficient approach in terms of the computational cost and therefore its suitability for real-time applications.

## 2. Kinematics Modeling Robot and its Odometery

The state of robot can be modeled as $(x, y, \theta)$ that $(x, y)$ are the Cartesian coordinates and $\theta$ is the orientation respectively to global environment. The kinematics equations for the mobile robot are in the following form [1-2] and [4]:

$$\begin{bmatrix} \dot{x} \\ \dot{y} \\ \dot{\phi} \end{bmatrix} = f(X) = \begin{bmatrix} (V + v_v)\cos(\phi + [\gamma + v_\gamma]) \\ (V + v_v)\sin(\phi + [\gamma + v_\gamma]) \\ \dfrac{(V + v_v)}{B}\sin(\gamma + v_\gamma) \end{bmatrix} \quad (1)$$

Where $B$ is the base line of the vehicle and $u = \begin{bmatrix} V & \gamma \end{bmatrix}^T$ is the control input consisting of a velocity input $V$ and a steer input $\gamma$, as shown in Fig.1.

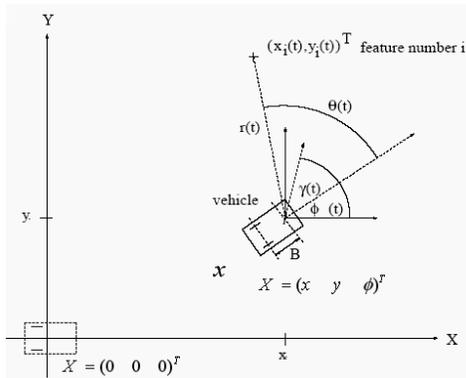

Fig.1 The robot and Feature

The process noise $v = \begin{bmatrix} v_v & v_\gamma \end{bmatrix}^T$ is assumed to be applied to the control input, $v_v$ to velocity input, and $v_\gamma$ to the steer angle input. The vehicle is assumed to be equipped with a sensor (range-laser finder) that provides a measurement of range $r_i$ and bearing $\theta_i$ to an observed feature $\rho_i$ relative to the vehicle as following:

$$\begin{bmatrix} r_i \\ \theta_i \end{bmatrix} = h(X) = \begin{bmatrix} \sqrt{(x - x_i)^2 + (y - y_i)^2} + \omega_r \\ \tan^{-1}\dfrac{y - y_i}{x - x_i} - \phi + \omega_\theta \end{bmatrix} \quad (2)$$

where $(x_i, y_i)$ is the position landmark in map and

$W = \begin{bmatrix} \omega_r & \omega_\theta \end{bmatrix}^T$ related to observation noise.

## 3. Extended Kalman Filter (EKF)

Kalman filter (KF) is widely used in studies of dynamic systems, analysis, estimation, prediction, processing and control. Kalman filter is an optimal solution for the discrete data linear filtering problem. KF is a set of mathematical equations which provide an efficient computational solution to sequential systems. The filter is very powerful in several aspects: It supports estimation of past, present, and future states (prediction), and it can do so even when the precise nature of the modeled system is unknown. The filter is derived by finding the estimator for a linear system, subject to additive white Gaussian noise. However, the real system is non-linear; Linearization using the approximation technique has been used to handle the non-linear system. This extension of the nonlinear system is called the Extended Kalman Filter (EKF). The general non-linear system and measurement form is as given by equations (3) and (4) as follows:

$$x_{k+1} = f(x_k, u_k) + w_k \quad (3)$$

$$z_k = h(x_k) + v_k \quad (4)$$

The system, measurement noises are assumed to be Gaussian with zero mean and are represented by their covariance matrices $Q_k$ and $R_k$:

$$E\{w_k\} = 0$$

$$E[w_k w_j^T] = \begin{cases} Q_k & k = j \\ 0 & k \neq j \end{cases} \quad (5)$$

$$E\{v_k\} = 0$$

$$E[v_k v_j^T] = \begin{cases} R_k & k = j \\ 0 & k \neq j \end{cases} \quad (6)$$

The Extended kalman filter algorithm has two groups of equations [14]:

1) The prediction equations:

The extended Kalman filter predicts the future state of system $\hat{x}_{k+1}^-$ based on the available system model $f(.)$ and projects ahead the state error covariance matrix $P_{k+1}^-$ using the time update equations:

$$\hat{x}_{k+1}^- = f(x_k, u_k) \quad (7)$$

$$P_{k+1}^- = \nabla f_k P_k \nabla f_k^T + G_u Q_k G_u^T \quad (8)$$

Where

$$\nabla f_k = \frac{\partial f}{\partial X} = \begin{bmatrix} \dfrac{\partial f_1}{\partial x} & \dfrac{\partial f_1}{\partial y} & \dfrac{\partial f_1}{\partial \phi} \\[2mm] \dfrac{\partial f_2}{\partial x} & \dfrac{\partial f_2}{\partial y} & \dfrac{\partial f_2}{\partial \phi} \\[2mm] \dfrac{\partial f_3}{\partial x} & \dfrac{\partial f_3}{\partial y} & \dfrac{\partial f_3}{\partial \phi} \end{bmatrix} \qquad (9)$$

$$G_u = \frac{\partial f}{\partial u} = \begin{bmatrix} \dfrac{\partial f_1}{\partial u_1} & \dfrac{\partial f_1}{\partial u_2} \\[2mm] \dfrac{\partial f_2}{\partial u_1} & \dfrac{\partial f_2}{\partial u_2} \\[2mm] \dfrac{\partial f_3}{\partial u_1} & \dfrac{\partial f_3}{\partial u_2} \end{bmatrix} \qquad (10)$$

2) Measurement updates equations

Once measurements $z_k$ become available the Kalman gain matrix $K_k$ is computed and used to incorporate the measurement into the state estimate $\hat{x}_k$. The state error covariance for the updated state estimate $P_k$ is also computed using the following measurement update equations:

$$K_k = P_k^- H_k^T (H_k P_k^- H_k^T + R_k)^{-1} \qquad (11)$$

$$\hat{x}_k = \hat{x}_k^- + K_k (z_k - h(\hat{x}_k^-)) \qquad (12)$$

$$P_k = (I - K_k H_k) P_k^- \qquad (13)$$

Where $I$ is an identity matrix and $H_k$ is following:

$$H_k = \frac{\partial h}{\partial X} \qquad (14)$$

In the above equations, $\hat{x}_k$ is an estimation of the system sate vector $x_k$ and $P_k$ is the covariance matrix corresponding to the state estimation error defined by

$$P_k = E\{(x_k - \hat{x}_k)(x_k - \hat{x}_k)^T\} \qquad (15)$$

The difference between the prediction and observed measurements is called the measurement innovation, or residual, generally denoted as $r_k$:

$$r_k = z_k - h(\hat{x}_k^-) \qquad (16)$$

The innovation represents the additional information variable to the filter in consequence to the observation $z_k$. For an optimal filter the innovation sequence is a sequence of independent Gaussian random variables.

## 4. Localization Based on EKF

We assume that robot knows map of environment. The EKF estimates a robot pose given a map of the environment and range-bearing of landmarks measurements. For this purpose, the data provided by odometric, map and laser range-finder are fused together by means of an EKF. To design EKF, The continuous time model formulated must be reformulated in the discrete time. The discrete time kinematics model is:

$$\begin{bmatrix} x(k+1) \\ y(k+1) \\ \phi(k+1) \end{bmatrix} = \begin{bmatrix} x(k) + T(V(k) + v_v(k))\cos(\phi(k) + [\gamma(k) + v_\gamma(k)]) \\ y(k) + T(V(k) + v_v(k))\sin(\phi(k) + [\gamma(k) + v_\gamma(k)]) \\ \phi(k) + T\dfrac{(V(k) + v_v(k))}{B}\sin(\gamma(k) + v_\gamma(k)) \end{bmatrix}$$

$$(17)$$

And the discrete time observation model is:

$$\begin{bmatrix} r_i(k) \\ \theta_i(k) \end{bmatrix} = \begin{bmatrix} \sqrt{(x(k) - x_i(k))^2 + (y(k) - y_i(k))^2} + \omega_r(k) \\ \tan^{-1}\dfrac{y(k) - y_i(k)}{x(k) - x_i(k)} - \phi(k) + \omega_\theta(k) \end{bmatrix}$$

$$(18)$$

Fig.2 briefly describes the cyclic localization procedure during the localization procedure using EKF.

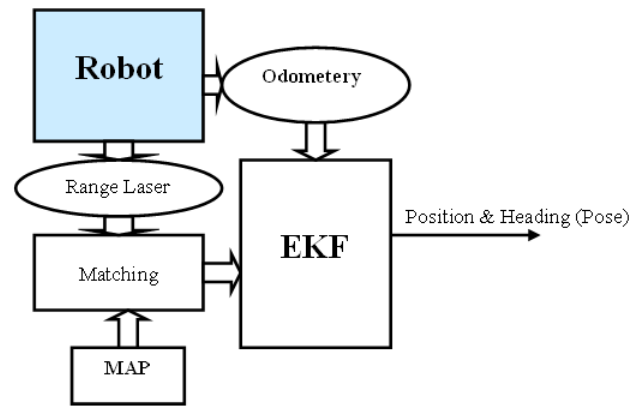

Fig.2 Localization algorithm based on EKF

The algorithm consists of such steps as position prediction, observation, measurement prediction, matching and estimation. The detailed discussion of these steps will be followed.

Step1) Initialization

Initialize the state vector $\hat{x}_0$ and covariance matrix $P_0$ of the mobile robot

Step 2) Robot Position Prediction

The robot position at time step k+1 is predicted based on its old localization (at time step k) and its movement due the control input $u_k$:

$$\hat{x}_{k+1}^- = f(x_k, u_k) \qquad (19)$$

$$P_{k+1}^- = \nabla f_k P_k \nabla f_k^T + G_u Q_k G_u^T \qquad (20)$$

Where

$$\nabla f_k = \frac{\partial f}{\partial X} \qquad G_u = \frac{\partial f}{\partial u} \qquad (21)$$

Step3) Measurement prediction

By using the predicted robot position $\hat{x}_{k+1}^-$ and the current map we can generate the predicted measurement $\hat{z}_k$ according to equation (22).

$$\hat{z}_k = \begin{bmatrix} \sqrt{(\hat{x} - x_i)^2 + (\hat{y} - y_i)^2} \\ \tan^{-1} \dfrac{\hat{y} - y_i}{\hat{x} - x_i} - \hat{\phi} \end{bmatrix} \qquad (22)$$

The error between the actual measurement $z_k$ and the predicted measurement based on estimation of the state is:

$$v_k = z_k - \hat{z}_k \qquad (23)$$

Where $v_k$ is innovation sequence (or residual) with the covariance:

$$S_k = \nabla h_k P_k^- \nabla h_k^T + R_k \qquad (24)$$

Where

$$\nabla h_k = \frac{\partial h}{\partial X} = \begin{bmatrix} \dfrac{\partial h_1}{\partial x} & \dfrac{\partial h_1}{\partial y} & \dfrac{\partial h_1}{\partial \phi} \\ \dfrac{\partial h_2}{\partial x} & \dfrac{\partial h_2}{\partial y} & \dfrac{\partial h_2}{\partial \phi} \end{bmatrix} \qquad (25)$$

Step 4) Matching

The goal of the matching procedure is to produce an assignment from measurements to the landmarks (stored in the map). If the measurement result satisfies the following inequality, it is thought to be eligible, otherwise, it is not, and it will be abnegated.

$$v_k S_k v_k^T \leq G \qquad (26)$$

Step 5) Estimation

$$\hat{x}_k = \hat{x}_k^- + K_k (z_k - h(\hat{x}_k^-)) \qquad (27)$$

Where $K_k$ is gain the kalman gain:

$$K_k = P_k^- \nabla h_k^T (\nabla h_k P_k^- \nabla h_k^T + R_k)^{-1} \qquad (28)$$

The new state covariance matrix is:

$$P_k = (I - K_k \nabla h_k) P_k^- \qquad (29)$$

Step 6) Return to step 2

# 5. Localization Based on Adaptive Neuro-Fuzzy EKF (ANFEKF)

As stated earlier, localization based on EKF assumes complete a priori knowledge of the process and measurement noise statistics; matrices $Q_k$ and $R_k$. However, in most application these matrices are unknown. An incorrect a prior knowledge of $Q_k$ and $R_k$ may lead to performance degradation [12] and it can even lead to practical divergence [13]. One of the effective ways to overcome the above mentioned weakness is to use an adaptive algorithm. Two major approaches that have proposed for adaptive EKF are Multiple Model Adaptive Estimation (MMAE) and Innovation Adaptive Estimation (IAE) [12]. In this paper IAE adaptive scheme of the EKF coupled with ANFIS to adjust the matrix $Q_k$ and $R_k$ is purposed. The ANFIS is used to adjust the EKF and is prevented the filter from divergence.

## 5.1 Localization based on ANFEKF ($Q_k$ is fixed)

The covariance matrix $R_k$ represents the accuracy of measurement instrument. Assuming that the noise covariance $Q_k$ is completely known, an algorithm to estimate the measurement noise covariance $R_k$ can be derived. In this case, an innovation based adaptive estimation (IAE) algorithm to adapt the measurement noise covariance matrix $R_k$ is derived. In particular, the technique known as covariance matching is used. The basic idea behind this technique is to make the actual value of the covariance of the residual to be consistent with its theoretical value [12]. The innovation sequence $r_k$ has a theoretical covariance $S_k$ that is obtained from EKF algorithm. The actual residual covariance $\hat{C}_k$ can be approximated by its sample covariance, through averaging inside a moving window of size N as the following:

$$\hat{C}_k = \frac{1}{N} \sum_{i=k-N+1}^{k} (r_i^T r_i) \qquad (30)$$

Where $i_0$ is first sample inside the estimation window. If the actual value of covariance $\hat{C}_k$ has discrepancies with its theoretical value, then the diagonal elements of $R_k$ based on the size of this discrepancy can be adjusted. The objective of these adjustments is to correct this mismatch as far as possible. The size of the mentioned discrepancy is given by a variable called the degree of mismatch ($DOM_k$), defined as

$$DOM_k = S_k - \hat{C}_k \qquad (31)$$

The basic idea used by an ANFIS, to adapt the matrix $R_k$ is as follows:

From equation (24) an increment in $R_k$ will increase $S_k$ and vice versa. Thus, $R_k$ can be used to vary $S_k$ in accordance with the value of $DOM_k$ in order to reduce the discrepancies between $S_k$ and $\hat{C}_k$. The adaptation of the $(i,i)$ element of $R_k$ is made in accordance with the $(i,i)$ element of $DOM_k$. The general rules of adaptation are as following:

If $DOM_k(i,i) \cong 0$ then maintain $R_k$ unchanged

If $DOM_k(i,i) > 0$ then decrease $R_k$

If $DOM_k(i,i) \cong 0$ then increase $R_k$

In this paper IAE adaptive scheme of the EKF coupled with adaptive Neuro-fuzzy inference system (ANFIS) is presented to adjust $R$.

### 5.1.1 The ANFIS Architecture ($Q_k$ is fixed)

The ANFIS model has been considered as two-input-single-output system. The inputs ANFIS are $DOM_k$ and $DeltaDOM_k$. Here, $DeltaDOM_k$ is defined as following:

$$DeltaDOM_k = DOM_k - DOM_{k-1} \qquad (32)$$

Fig.3 and Fig.4 present membership functions for $DOM_k(i,i)$ and $DeltaDOM_k$ as shown.

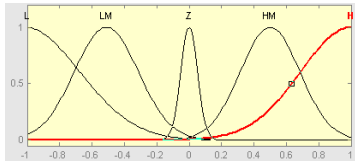

Fig.3 Membership function $DOM_k$

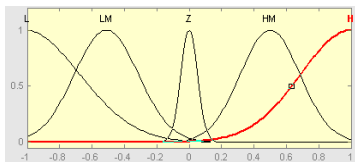

Fig .4 Membership function $DeltaDOM_k$

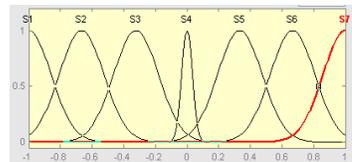

Fig .5.Membership function $AjdR_k$

In addition, adjustments of $R_k$ is performed using the following relation

$$R_k = R_k + \Delta R_k \qquad (33)$$

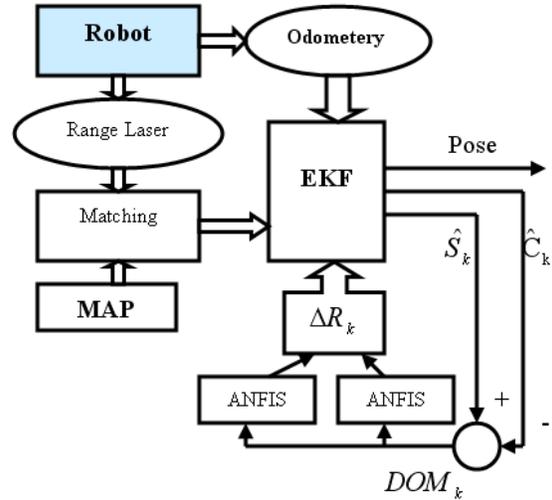

Fig.6 Localization Based on ANFEKF ($Q_k$ Fixed)

where $\Delta R_k$ is ANFIS output and membership function of $\Delta R_k$ is shown in fig.5. As size $DOM_k$ and $R$ is two, two system ANFIS to adjust EKF is used as shown in Fig.6. The following structure which is a five layers network is proposed in fig .7.

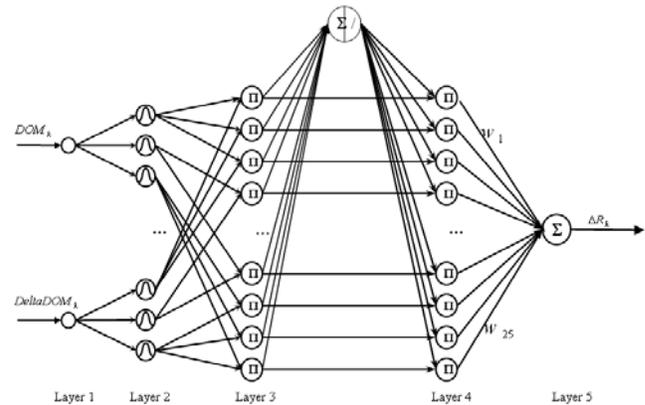

Fig .7 The ANFIS Architecture ($Q_k$ is fixed)

Let $u_i^l$ and $o_i^l$ denote the input to output from the $i$ th node of the $l$ th layer, respectively. To provide a clear understanding of an ANFIS, the function of layer 1 to layer 5 are defined as follows:

**Layer 1:** The node in this layer only transmits input values to the next layer directly, i.e.

$$o_i^1 = u_i^1 \qquad (34)$$

**Layer2:** In this layer, each node only performs a membership function. Here, the input variable is fuzzified by using five membership functions (MFs). The output of the $i$ th MF is given as:

$$o_{ij}^2 = \mu_{ij}(u^2) = \exp\left\{-\frac{(u_{ij}^2 - m_{ij})^2}{(\delta_{ij})^2}\right\} \qquad (35)$$

Where $m_{ij}$ and $\delta_{ij}$ are the mean and with of the Gaussian membership function, respectively. The subscript $ij$ indicates the $jth$ term of the $ith$ input. Each node in this layer has two adjustable parameters: $m_{ij}$ and $\delta_{ij}$

Layer3: The nodes in this layer are rule nodes. The rule node performs a fuzzy and operation (or product inference) for calculating the firing strength.

$$o_i^3 = \prod_l u_l^3 \qquad (36)$$

**Layer4:** The node in this layer performs the normalization of firing strengths from layer 3,

$$o_l^4 = \frac{u_l^4}{\sum_{l=1}^9 u_l^4} \qquad (37)$$

**Layer5:** This layer is the output layer. The link weights in this layer represent the singleton constituents ($W_i$) of the output variable. The output node integrates all the normalization firing strength from layer 4 with the corresponding singleton constituents and acts as defuzzfier,

$$\Delta R_i = \sum_{l=1}^{25} u_l^5 w_l \qquad (38)$$

The fuzzy rules which complete the ANFIS rule base are as table.1.

Table.1: Rule Table

| $DeltaDOM_k$ $DOM_k$ | L | LM | Z | HM | H |
|---|---|---|---|---|---|
| L | S7 | S7 | S6 | S5 | S4 |
| LM | S7 | S6 | S5 | S4 | S3 |
| Z | S6 | S5 | S4 | S3 | S2 |
| HM | S5 | S4 | S3 | S2 | S1 |
| H | S4 | S3 | S2 | S1 | S1 |

### 5.1.2 Learning Algorithm

The aim of the training algorithm is to adjust the network weights through the minimization of following cast function:

$$E = \frac{1}{2}e_k^2 \qquad (39)$$

Where

$$e_k = S_k - \hat{C}_k \qquad (40)$$

By using the back propagation (BP) learning algorithm, the weighting vector of the ANFIS is adjusted such that the error defined in (39) is less than a desired threshold value after a given number of training cycles. The well-known BP algorithm may be written as:

$$W(k+1) = W(k) + \eta(-\frac{\partial E(k)}{\partial W(k)}) \qquad (41)$$

Here $\eta$ and $W$ represent the learning rate and tuning parameter of ANFIS respectively. Let $W = [m,\sigma,w]^T$ be the weighting vector of ANFIS. The gradient of $E$ with respect to an arbitrary weighting vector $W$ is as the following:

$$\frac{\partial E(k)}{\partial W(k)} = -e(k)\left(\frac{\partial \Delta R(k)}{\partial W(k)}\right) \qquad (42)$$

By recursive applications of chain rule, the error term for each layer is first calculated, and then the parameters in the corresponding layers are adjusted.

### 5.2 Localization based on ANFEKF ($R_k$ Fixed)

Assuming that the noise covariance matrix $R_k$ is completely known an algorithm to estimate matrix $Q_k$ can be derived. The idea behind the process of adaptation of $Q_k$ is as follows:Equation (8) can be rewritten as

$$S_k = \nabla h_k (\nabla f_k P_k \nabla f_k^T + G_u Q_k G_u^T) \nabla h_k^T + R_k \qquad (43)$$

It may be deduced from equation (43) that a variation in $Q_k$ will affect the value of $S_k$. If $Q_k$ is increased, then $S_k$ is increased, and vice versa. Thus, if a mismatch between $S_k$ and $\hat{C}_k$ is observed then a correction can be made through augmenting or diminishing the value of $Q_k$.The tree general adaptation rules are defined as following

1. If $DOM_k(1,1)$ is L and $DOM_k(2,2)$ is L hen $\Delta Q_k$ is H

2. If $DOM_k(1,1)$ is Z and $DOM_k(2,2)$ is Z then $\Delta Q_k$ is Z

3. If $DOM(1,1)$ is H $DOM(2,2)$ is H then $\Delta Q_k$ is L

Then $Q_k$ is adapted in this way

$$Q_k = Q_k \Delta Q_k \qquad (44)$$

Where $\Delta Q_k$ is the ANFIS output and $DOM_k(1,1)$ and $DOM_k(2,2)$ are ANFIS input.

The ANFIS model has been considered as a two-input-single-output system as previous section. Fig.8 shows the block diagram of Localization based on ANFEKF while $R_k$ is Fixed.

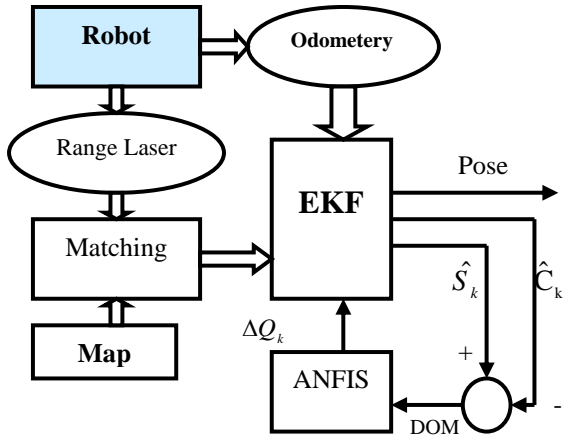

Fig.8  Localization Based on ANFEKF ($R_k$ Fixed)

## 6. Implementation and Results

Experiments have been carried out to evaluate the performance of the proposed approach in comparison with classical method.

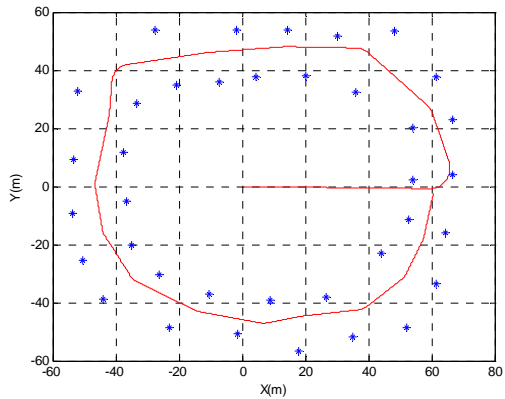

Fig.9 The experiment environment: The star point "*" denote the landmark

Fig.9 shows the robot trajectory and landmark location. The star points (*) depict location of the Landmarks that are known. The initial position of the robot is assumed to be $x_0 = 0$. The robot moves at a speed 3m/s and with a maximum steering angle 30 deg. Also, the robot has 4 meters wheel base and is equipped with a range-bearing sensor with a maximum range of 20 meters and a 180 degrees frontal field-of-view. The control noise is $\sigma_v = 0.3$ m/s and $\sigma_\gamma = 3^o$. A control frequency is 40 HZ and observation scans are obtained very 5 HZ. The measurement noise is 0.1 m in range and $1^o$ in bearing.

The performance of the proposed method is compared with localization based on EKF where matrices $Q_k$ and $R_k$ are kept static throughout the experiment The proposed method starts with the same $Q_k$ and $R_k$ matrices, but it keeps adapting the $Q_k$ and $R_k$ matrices according to the proposed scheme. The localization based on is known as a good choice when the associated statistical models are well known. Fig.10 shows performance of Localization based EKF and proposed method in this situation. It is observed that performance both algorithms are almost same.

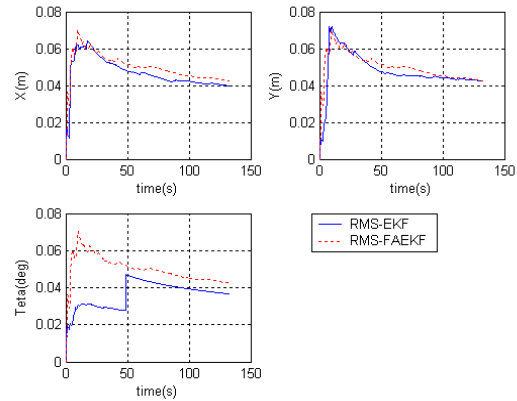

Fig.10 The figure shows the RMS error in localization Based on AFEKF and EKF (the results obtain over 50 Monte Carlo runs). In this experiment, measurement noise is $\sigma_r = 0.1$, $\sigma_\theta = 1.0$ and control noise is $\sigma_v = 0.3$ m/s, $\sigma_\gamma = 3^o$.

However, the performance of localization based of EKF degrades when the knowledge of such statistics is inappropriate. For this purpose, we first consider the situation where the sensor statistics are set wrongly as: $\sigma_r = 2.0$ and $\sigma_\theta = 0.1$ and the noise covariance $Q_k$ is completely known. The proposed algorithm starts with a wrongly know statistics and then it adapts the $R_k$ matrix, online, on the basis of ANFIS attempts to minimize the mismatch between the theoretical and actual values of the innovation sequence. The free parameters of ANFIS are automatically learned by SD during training. Fig.11 shows the root mean square error (RMSE) of localization based on EKF and FAEKF for 25 different runs. It is observed that state estimates from the FAEK are more accurate than the EKF.

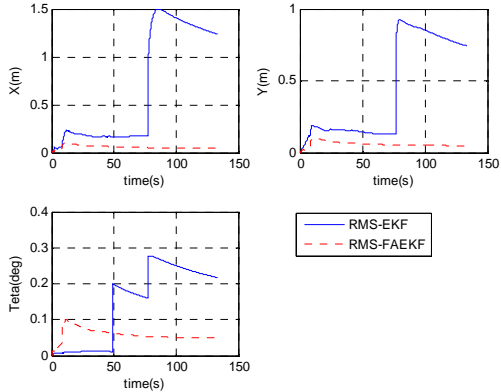

Fig.11 The figure shows the RMS error in localization Based on AFEKF and EKF (the results obtain over 50 Monte Carlo runs). In this experiment, measurement noise is wrongly considered as $\sigma_r = 2$, $\sigma_\theta = 0.1$ and control noise is truly considered as $\sigma_v = 0.3$ m/s , $\sigma_\gamma = 3^o$ .

Now, we consider the situation where the uncertainties in control inputs are wrongly considered as: $\sigma_\varpi = 0.03$ m/s , $\sigma_\theta = 0.5$ deg and measurement covariance $R_k$ is completely known. The proposed algorithm starts with a wrongly know statistics and then adapt the $Q_k$ matrix, online, on the basis of ANFIS attempts to minimize the mismatch between the theoretical and actual values of the innovation sequence. Fig.12 shows RMSE for this situation. Such as previous situation, it is observed that localization based on FAEKF is more accurate than localization based on EKF.

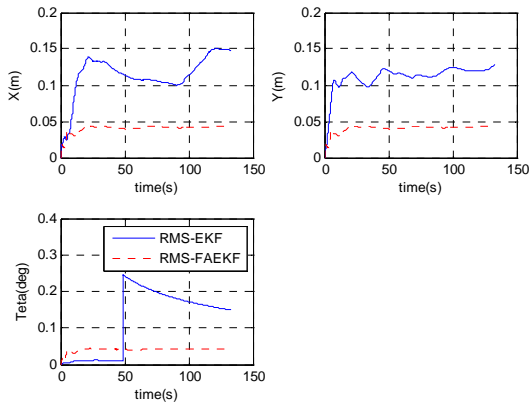

Fig.12 The figure shows the RMS error in localization Based on AFEKF and EKF (the results obtain over 50 Monte Carlo runs). In this experiment, measurement noise is truly considered as $\sigma_r = 2$ , $\sigma_\theta = 0.1$ and control noise is wrongly considered as $\sigma_v = 0.03$ m/s , $\sigma_\gamma = 0.5^o$

Finally, the consistency of both localizations Based on AFEKF and EKF are compared. In this regard, we have considered the situation where the sensor statistics are set wrongly as: $\sigma_r = 2.0$ and $\sigma_\theta = 0.5$ and the noise covariance $Q_k$ is completely known. To verify the consistency of both algorithms, average Normalized Estimation Error Squared (NEES) is used as a measure factor. For an available ground truth $x_k$ and an estimated mean and covariance $\left\{ \hat{x}, \hat{P} \right\}$ , we can use NEES to characterize the filter performance:

$$\varepsilon_k = (x_x - \hat{x}_k)^T P_k^{-1} (x_x - \hat{x}_k) . \tag{45}$$

Consistency is evaluated by performing multiple Monte Carlo runs and computing the average NEES. Given $N$ runs, the average NEES is computed as

$$\overline{\varepsilon}_k = \frac{1}{N} \sum_{i=1}^{N} \varepsilon_{ik} \tag{46}$$

Given the hypothesis of a consistent linear-Gaussian filter, $N \overline{\varepsilon}_k$ has a $\chi^2$ density with $N \dim(x_k)$ degrees of freedom [17]. Thus, for the 3-dimensional vehicle pose, with Twenty Monte Carlo simulations, the two sided 95% probability concentration region for $\overline{\varepsilon}_k$ is bounded by interval [2.02, 4.17].

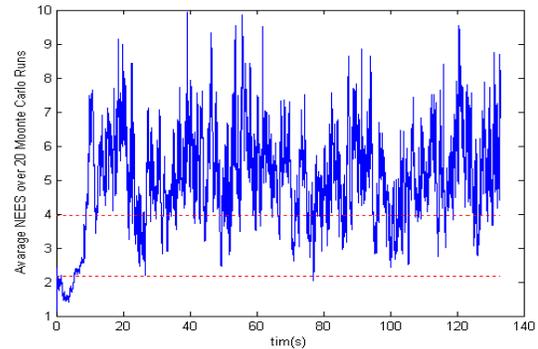

Fig.13 Consistency of localization based on EKF

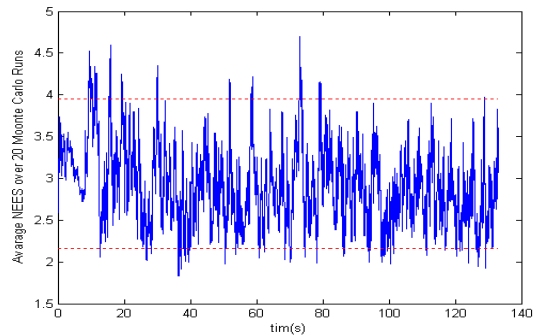

Fig.14 Consistency of localization based on AFEKF

Fig.13 and Fig.14 show that the consistency of localization Based on AFEKF is more than that of localization Based on EKF.

## 7. Conclusion

The preset paper has proposed the development of a adaptive Neuro-Fuzzy inference system for localization based EKF when a priori knowledge is incorrect. The EKF is known as a choice for localization when a priori knowledge is well known. However, incorrect knowledge of these statistics can cause significant degradation in performance. The present scheme proposes to start the system with a wrongly known statistics and adapt the matrices $R_k$ and $Q_k$, online, on the basis of ANFIS that attempt to minimize the mismatch between the theoretical and actual values of the innovation sequence. The free parameters of ANFIS are automatically learned by employing the steepest gradient descent during the training. Main advantage of proposed method is that consistency of this approach is more than localization based EKF. This is because that theoretical value of the innovation sequence is match with its actual value in proposed method. The two experiments in simulation have been carried out to study of performance the proposed method. The simulation results are shown that localization based on AFEKF is more accurate than localization based on EKF.

## References

[1] F.Kong, Y.Chen, J.Xie, Gang, "Mobile robot localization based on extended kalman filter", Proceedings of the 6th World Congress on Intelligent Control and Automation, June 21-23, 2006.

[2] J.Kim, Y.Kim, and S.Kim, "An accurate localization for mobile robot using extended kalman filter and sensor fusion", Proceeding of the 2008 International Joint Conference on Neural Networks (IJCNN 2008).

[3] Tran Huu Cong, Young Joong Kim and Myo-Taeg Lim, "Hybrid Extended Kalman Filter-based Localization with a Highly Accurate Odometry Model of a Mobile Robot", International Conference on Control, Automation and Systems 2008

[4] Sangjoo Kwon, Kwang Woong Yang and Sangdeok Park, "An Effective Kalman filter Localization Method for mobile Robots", Proceeding of the 2006 IEEE/RSJ, International Conference on Intelligent Robots and Systems.

[5] G.Reina, A.Vargas, KNagatani and K.Yoshida "Adaptive Kalman Filtering for GPS-based Mobile Robot Localization", in Proceedings of the 2007 IEEE, International Workshop on Safety, Security and Rescue Robotics.

[6] M.Betke and L.Gurvits, "Mobile robot localization using landmarks",IEEE Trans on Robotics And Automation, Vol. 13, No. 2, APRIL 1997.

[7] I.Roumeliotis and A.Bekey, "An extended kalman filter for frequent local and infrequent global sensor data fusion". in SPIE International Symposium on Intelligent Systems and Advanced Manufacturing,1997.

[8] W.Jin, X.Zhan, A modified kalman filtering via fuzzy logic system for ARVs Localization", Proceeding of the 2007 IEEE, International Conference on Mechatronics and Automation.

[9] L.Jetto, S.Longhi, "Development and experimental validation of an adaptive extended kalman filter for the localization of mobile robots", IEEE Trans on robotics and automation, vol.15, No.2, April 1999.

[10] E.E.EI Madbouly, A. E. Abdalla, Gh. M. EI Banby, "Fuzzy adaptive kalman filter for multi sensor system", 2009 IEEE

[11] Wei Qin, Weizheng Yuan, Honglong Chang, Liang Xue, Guangmin Yuan, "Fuzzy Adaptive Extended Kalman Filter for miniature Attitude and Heading Reference System," 4th IEEE International Conference on Nano/Micro Engineered and Molecular Systems, 2009.

[12] R.K.Mehra, "On the identification of variances and adaptive kalman filtering", IEEE Trans, Autom. Control, Vol. AC-15, No. 2, pp. 175–184, Apr. 1970.

[13] R.J.Fitzgerald, "Divergence of the kalman filter", IEEE Trans. Autom. Control, Vol. AC-16, No. 6, pp. 736–747, Dec. 1971.

[14] M.S.Grewal and A.P.Andrews, "Kalman filtering: Theory and practice", Prentice-HALL, 1993.

[15]Chatterjee, A. Matsuno, F. "A Neuro-Fuzzy Assisted Extended Kalman Filter-Based Approach for Simultaneous Localization and Mapping (SLAM) Problems", IEEE Transactions on Fuzzy Systems, Oct. 2007, 15: 5, 984-997

[16] Wu, Z. Q. and Harris, C. J. Adaptive Neurofuzzy Kalman Filter. In: FUZZ-IEEE '96 - Proceedings of the fifth IEEE International Conference on Fuzzy Systems. pp. 1344-1350.

[17] Y. Bar-Shalom, X.R. Li, and T. Kirubarajan. Estimation with Applications to Tracking and Navigation. John Wiley and Sons, 2001.

[18] J. Z. Sasiadek, Q.Wang, and M. B. Zeremba, "Fuzzy adaptive Kalman filtering for INS/GPS data fusion," in proc. 15th Int. Symp. Intell. Control,Patras, Greece, Jul 2000.

[19] D. Loebis, R. Sutton, J. Chudley, and W. Naeem, Adaptive tuning of a Kalman filter via fuzzy logic for an intelligent auv navigation system," Control Eng. Pract., vol. 12, pp. 1531–1539, 2004.

[20] L. Jetto, S. Longhi, and D. Vitali, .Localization of a wheeled mobile robot by sensor data fusion based on a fuzzy logic adapted Kalman filter ,. Control Engineering Practice, vol. 7, pp. 763-771, 1999.

**Ramazan Havangi** received the M.S. degree in Electrical Engineering from K.N.T.U University, Tehran, Iran, in 2004; He is currently working toward the Ph.D. degree in K.N.T.U University. His current research interests include Inertial Navigation, Integrated Navigation, Estimation and Filtering, Evolutionary Filtering, Simultaneous Localization and Mapping, Fuzzy, Neural Network, and Soft Computing.

**Mohammad Ali Nekoui** is assistant professor at Department of Control, Faculty of Electrical Engineering K.N.T.U University. His current research interests include Optimal Control Theory, Convex Optimization, Estimation and Filtering, simultaneous localization and mapping, Evolutionary Filtering, Simultaneous Localization and Mapping.

**Mohammad Teshnehlab** is professor at Department of Control, Faculty of Electrical Engineering, K.N.T.U University. His current research interests include Fuzzy, Neural Network, Soft Computing, Evolutionary Filtering, and Simultaneous Localization and Mapping.